# *Affine Tensors in Mechanics of Freely Falling Particles and Rigid Bodies*

GÉRY DE SAXCÉ[1], CLAUDE VALLÉE[2]

[1] Laboratoire de Mécanique de Lille, UMR CNRS 8107,
Université de Lille-1, Cité scientifique, F59655 Villeneuve d'Ascq, France
phone : +33 320 336 020, fax : +33 320 337 153, e-mail: gery.desaxce@univ-lille1.fr

[2] Laboratoire de Mécanique des Solides, UMR CNRS 6610,
SP2MI, BP 30179, F86962 Futuroscope-Chasseneuil, France
phone : +33 549 496 792, fax : +33 549 496 504, e-mail: vallee@lms.univ-poitiers.fr

This version: 2006-12-04

## Abstract

In a previous paper, we proposed an approach for the dynamics of 3D bodies and shells based on the use of affine tensors. This new theoretical frame is very large and the applications are not limited to the mechanics of continua. In the present paper, we show how it can be also applied to the description of the mechanics of freely falling particles and rigid bodies.

The mass, the linear and angular momenta are structured as a single object called torsor. Excluding all metric notions, we define the torsors as skew-symmetric bilinear mappings operating on the linear space of the affine functions. Torsors are a particular family of affine tensors.

On this ground, we define an intrinsic differential operator called the affine covariant derivative. Next, we claim that the torsor characterizing the behavior of a freely falling particle is affine covariant derivative free, that allows recovering both laws of linear and angular momentum. Finally, it is shown how the motion of rigid bodies can be describe within this frame.

# 1. Introduction

The moment of a force, due to Archimedes, is a fundamental concept of the mechanical science. Its modelling by means of standard mathematical tools is well known. In the modern literature, it sometimes appears in the axiomatic form of the concept of torsor (Pérès 1953), an object composed of a vector and a moment, endowed with the property of equiprojectivity and obeying a specific transport law. Although this last one invokes a translation of the origin, a very few interest has be taken in wondering about the affine nature of this object. These elementary notions can be presented with a minimal background of vector calculus. At a higher mathematical level, another not less overlooked keystone of the Mechanics is the concept of continuous medium, especially organised around the tensorial calculus which stems from Cauchy's works about the stresses (Cauchy 1923 and 1927). The general rules of this calculus were introduced by Ricci-Curbasto and Levi-Civita 1901. They are concerned by the tensors that we shall call 'linear tensors' insofar as their components are modified by means of linear frame changes, then of regular linear transformations, elements of the linear group. The use of moving frames allows determining these objects in a covariant way thanks to a connection, known by its Christoffel's symbols.

It was É. Cartan who pointed out the fact that the linear moving frames could be replaced by affine moving frames, introducing so the affine connexions (É. Cartan 1923 and 1924). His successors only remember the concepts of principal bundle and connection associated to some group: the linear group, the affine group, the projective group and so on. It is the applications to the orthogonal group which above all will hold the attention on account of the Riemannian geometry and the Euclidean tensors. The interest that Cartan originally took in the connections of the affine group became of secondary importance. Perhaps only remains the name of 'affine connection', while oddly used for any group, even if it is not affine. Certainly one can find in the continuous medium approach a unifying tool of the Mechanics, even if the dynamics of the material particles and the rigid bodies remain on the fringe and if the torsor —so essential to the Mechanics— seems to escape from any attempt of getting it into the mould of the tensorial calculus. The contemporaries rather will find responses to this worry of unifying and structuring the Mechanics in the method of virtual powers or works, initiated by Lagrange, and the variational techniques. Without denying the power of these tools, their abstract character and the traps of the calculus of variations must not be underestimated yet.

More recently, Souriau proposes to revisit the Mechanics emphasizing its affine nature (Souriau 1997-a). It is this viewpoint that we will adopt here, starting from a generalisation of the concept of torsor under the form of an affine object (de Saxcé et al. 2003). It allows structuring the Mechanics thanks to a unique principle which, by declining it for each kind of continuous medium, provides the classical equations of the statics and dynamics.

Our starting point is closely related to Souriau approach on the ground of two key ideas: a new definition of torsors and the crucial part played by the affine group of $\mathbb{R}^n$. This group forwards on a manifold an intentionally poor geometrical structure. Indeed, this choice is guided by the fact that it contains both Galilei and Poincaré groups (Souriau 1997-b), which allows involving the classical and relativistic mechanics at one go. This viewpoint implies that we do not use the trick of the Riemannian structure. In particular, the linear tangent space cannot be identified to its dual one and tensorial indices may be neither lowered nor raised.

A class of tensors corresponds to each group. The components of these tensors are transformed according to the action of the considered group. The standard tensors discussed in the literature are those of the linear group of $\mathbb{R}^n$. We will call them linear tensors. A fruitful standpoint consists in considering the class of the affine tensors, corresponding to the affine group (de Saxcé et al. 2003).

To each group is associated a family of connections allowing to define covariant derivatives for the corresponding classes of tensors. The connections of the linear group are known through Christoffel's coefficients. They represent, as usual, infinitesimal motions of the local basis. From a physical viewpoint, these coefficients are force fields such as gravity or Coriolis' force. To construct the connection of the affine group, we need Christoffel's coefficients stemming from the linear group and additional ones describing infinitesimal motions of the origin of the affine space associated with the linear tangent space. On this ground, we construct the affine covariant divergence of torsors. The concept of torsor was successfully applied to the dynamics of three-dimensional bodies and shells (de Saxcé et al. 2003). We claim that the torsor field representing the behavior of these continua is affine divergence free, which allow recovering the motion equations. The aim of the present paper is to show that this concept is also relevant to describe in a simple and elegant manner the dynamics of freely falling particles.

In the classical Mechanics, a free particle moves in straight line with uniform velocity. In order to give a more detailed definition of this « uniform straight motion », one considers a particle of velocity $v$, constant mass $m$ with position $r$ at time $t$ with respect to a given origin of the space, and position $r_0$ at time $0$. The trajectory is given by: $r = r_0 + v\,t$. The motion is classically described by the linear and angular momenta:
$$p = mv, \qquad l = m\,r \times v. \qquad (1.1)$$
Let us consider now the position $r' = r + k'$ of the particle about another space origin. The angular momentum is changed according to the following *momentum transport* law:
$$l' = l + k' \times p, \qquad (1.2)$$
while the linear momentum is unchanged (for instance, see J.-L. Synge 1960). A wide set of experiments in Mechanics can be clearly interpreted by means of this simple rule which must be considered as an experimental evidence.

The present paper is a detailed and extended version of de Saxcé et al. 2005-a. The organisation of the paper is as follows. In Section 2, we briefly define the class of affine tensors and, in particular, the torsors. Using heuristic arguments, we show the pertinence of using affine tensors by deducing the momentum transport law (1.2) from the tensorial rules of torsors. In Section 3, we restrict the analysis to the classical mechanics by considering Galilei group. A pervasive study of the invariants of the Galilean torsors allows revealing the physical interpretation of the torsor components. In Section 4, we present a generalization of the usual "Tensorial Analysis" in order to differentiate in an intrinsic way the fields of affine tensors. In Section 5, we claim that the affine covariant derivative of the torsor is null. We deduce the motion equations, including the well-known theorem of the angular momentum but recast in a more general way which highlights the covariant form of this law with respect to any moving origin. In Section 6, the rigid body is shown to be equivalent to a particle with spin and Poinsot formulation is recovered.

# 2. Affine tensors

**Linear tensors**. They are widely used now in various branches of the Mechanics and the Physics. Many classical reference books are devoted to them. The aim of this section is only to define the notations used in the sequel. Let $\mathscr{T}$ be a linear space (or vector space) of dimension $n$, and $(\vec{e}_\alpha)$ be a basis of $\mathscr{T}$. The associated co-basis $(\vec{e}^\alpha)$ is such that: $\vec{e}^\alpha(\vec{e}_\beta) = \delta^\alpha_\beta$. A new basis: $\vec{e}_{\alpha'} = P^\beta_{\alpha'} \vec{e}_\beta$ can be defined through the transformation matrix $P = (P^\beta_{\alpha'})$. We denote the inverse matrix $P^{-1}$. The simplest types of linear tensors are the vectors and the covectors (or linear forms). Some typical tensors are recalled in Table 1.

*Table 1*. Linear tensors.

| Tensor type | space $\boldsymbol{T}^{q\ contravariant}_{p\ covariant}\}index$ | abstract notations | transformation laws of the components |
|---|---|---|---|
| vector | $\mathscr{T} = \boldsymbol{T}^1_0$ | $\vec{V} = V^\alpha \vec{e}_\alpha$ | $V^{\alpha'} = (P^{-1})^{\alpha'}_\beta V^\beta$ |
| covector (linear form) $\mathscr{T} \xrightarrow{\bar{\Phi}\ linear} \mathbb{R}$ | (dual space) $\mathscr{T}^* = \boldsymbol{T}^0_1$ | $\bar{\Phi} = \Phi_\alpha \vec{e}^\alpha$ | $\Phi_{\alpha'} = \Phi_\beta P^\beta_{\alpha'}$ |
| tensor of contravariant order 2 $\mathscr{T}^* \times \mathscr{T}^* \xrightarrow{M\ bilinear} \mathbb{R}$ | $\boldsymbol{T}^2_0$ | $M = M^{\alpha\beta} \vec{e}_\alpha \otimes \vec{e}_\beta$ | $M^{\alpha'\beta'} = (P^{-1})^{\alpha'}_\mu (P^{-1})^{\beta'}_\nu M^{\mu\nu}$ |

**Generalizing the concept of tensor**. Now, the main features of the usual concept of tensor are discussed. Firstly, the tensors can be seen as multilinear mappings. In this case, they are represented using the abstract notation. Another viewpoint is considering an assignment of components to each frame. It leads to use the tensorial notation. The changing of frames obeys suitable transformation laws. Finally, if no reference to frame is done, the tensors can be seen as objects the components of which are changed according to prescribed transformation laws. For the usual tensors, the frames are the basis of a linear space and the transformations are the regular linear mappings from $\mathbb{R}^n$ into itself or, in other words, the $n \times n$ regular matrices $P$. The set of all such transformations is the linear group $\mathbb{GL}(n)$.

This discussion suggests generalizing the usual concept of tensor in the following way: *the tensors are objects the components of which are changed by a given group of transformations (not necessarily linear)*. Considering the linear group $\mathbb{GL}(n)$, we recover the class of the usual tensors. Nevertheless, other choices of transformation groups are possible. In many applications, people customarily handle the orthogonal group $\mathbb{O}(n)$, a subgroup of $\mathbb{GL}(n)$, that leads to the class of the Euclidean tensors. On the other hand, we could consider the affine group $\mathbb{GA}(n)$, an extension of $\mathbb{GL}(n)$ obtained by adding the translations.

Which tensor class is the most relevant? It is the experiment which must be our guide in this choice. *Therefore, the momentum transport rule* (1.2) *is able to explain experiments in Mechanics. This rule deals with the origin changing in an affine space. Hence, this suggests the natural tensors to describe the mechanical experiments are the affine tensors.*

The simplest types of affine tensors are the points of an affine space and the affine functions on this space with real values, corresponding respectively to the vectors and covectors (the simplest types of linear tensors).

**Points of an affine space**. To define an origin $Q$ of the affine space $A\mathcal{T}$ associated to $\mathcal{T}$, we can use the column vector $V_0$ collecting the components $V_0^\alpha$, in the basis $(\vec{e}_\alpha)$, of the vector $\overrightarrow{Q0}$ joining $Q$ to the zero of $\mathcal{T}$ considered as point of $A\mathcal{T}$. By the choice of this affine frame $r = (V_0, (\vec{e}_\alpha))$, any point $V$ of $A\mathcal{T}$ can be identified to the vector $\overrightarrow{QV} = V^\alpha \vec{e}_\alpha$. Now, let $r' = (V_0', (\vec{e}_{\alpha'}))$ be a new affine frame of origin $Q'$ and basis $(\vec{e}_{\alpha'})$. In the next, If $C'$ is the column vector collecting the components $C^{\alpha'}$ of the translation $\overrightarrow{Q'Q}$ in the new basis, the origin changing is obtained by the decomposition $\overrightarrow{Q'0} = \overrightarrow{Q'Q} + \overrightarrow{Q0}$ in the new basis, leading to the following relation:
$$V_0^{\alpha'} = C^{\alpha'} + (P^{-1})^{\alpha'}_\beta V_0^\beta . \tag{2.1}$$
The set of all the affine transformations $(C', P^{-1})$ is the affine group $\mathbb{GA}(n)$. Similarly, by decomposition of $\overrightarrow{Q'V} = \overrightarrow{Q'Q} + \overrightarrow{QV}$ in the new basis, we obtain the corresponding transformation law for the affine components $V^\beta$ of the point $V$:
$$V^{\alpha'} = C^{\alpha'} + (P^{-1})^{\alpha'}_\beta V^\beta .$$
In matrix notation, one reads:
$$V' = C' + P^{-1} V .$$
The frames such that $V_0 = 0$ can be identified to the basis of the associated linear space and play a particular role in the sequel. We call them *proper affine frames*.

**Affine functions**. Any affine mapping $\psi$ from $A\mathcal{T}$ into $\mathbb{R}$ is called an affine function of $A\mathcal{T}$. An affine mapping $\psi$ is represented in an affine frame $r$ by:
$$\psi(V) = \chi + \Phi_\alpha V^\alpha ,$$
where $\chi = \psi(Q)$ and $\Phi_\alpha$ are the components, in the co-basis $(\bar{e}^\alpha)$, of the unique covector $\bar{\Phi}$ associated to $\psi$. We will call $(\Phi_\alpha, \chi)$ the affine components of $\psi$. After a change of affine frame, they are given by:
$$\Phi_{\alpha'} = \Phi_\beta P^\beta_{\alpha'}, \qquad \chi' = \chi - \Phi_\beta P^\beta_{\alpha'} C^{\alpha'}, \tag{2.2}$$
as it can be easily verified. It is worthwhile to remark that the set $A^*\mathcal{T}$ of such functions is a linear space of dimension $(n+1)$.

**Torsors**. Of course, affine tensors of more complex type can be constructed, for instance the ones that we called torsors. Any bilinear skew-symmetric mapping $\mu$ from $A^*\mathcal{T} \times A^*\mathcal{T}$ into $\mathbb{R}$ is called a torsor. Hence:
$$\forall \psi, \hat{\psi} \in A^*\mathcal{T}, \quad \mu(\psi, \hat{\psi}) = -\mu(\hat{\psi}, \psi) .$$
With respect to the affine coframe $r$, the torsor $\mu$ is represented by:
$$\mu(\psi, \hat{\psi}) = \mu((\Phi_\alpha, \chi), (\hat{\Phi}_\beta, \hat{\chi})) = J^{\alpha\beta} \Phi_\alpha \Phi_\beta + T^\alpha (\chi \hat{\Phi}_\alpha - \hat{\chi} \Phi_\alpha) ,$$
with: $J^{\alpha\beta} = -J^{\beta\alpha}$.
The affine components of the torsor $\mu$ are $(T^\alpha, J^{\alpha\beta})$. Accounting for equation (2.2), it is easy to deduce the corresponding transformation laws:
$$T^{\alpha'} = (P^{-1})^{\alpha'}_\beta T^\beta , \qquad J^{\alpha'\beta'} = (P^{-1})^{\alpha'}_\mu (P^{-1})^{\beta'}_\nu J^{\mu\nu} + C^{\alpha'}((P^{-1})^{\beta'}_\mu T^\mu) - ((P^{-1})^{\alpha'}_\mu T^\mu) C^{\beta'} .$$
In matrix notation, one has equivalently:

$$T' = P^{-1} T, \qquad J' = P^{-1} J P^{-T} + C'(P^{-1}T)^T - (P^{-1}T)C'^T. \tag{2.3}$$

The torsors have a straightforward mechanical interpretation. Any event occurring at position $r$ and time $t$ can be represented by a point of the space-time of coordinates:

$$X = \begin{pmatrix} t \\ r \end{pmatrix} \in \mathbb{R}^4.$$

For convenience, we introduce the cross-product mapping $j(v)$ (sometimes also denoted $ad(v)$):

$$j(v) = \begin{pmatrix} 0 & -v^3 & v^2 \\ v^3 & 0 & -v^1 \\ -v^2 & v^1 & 0 \end{pmatrix},$$

such that: $v \times w = j(v) w$. Next, we define the $T^\alpha$ components as:

$$T = \begin{pmatrix} m \\ p \end{pmatrix}, \tag{2.4}$$

where $m$ is the mass and $p$ is the linear momentum. Next, we put:

$$J = \begin{pmatrix} 0 & -q^T \\ q & -j(l) \end{pmatrix}, \tag{2.5}$$

where $l$ is the angular momentum and the physical meaning of $q \in \mathbb{R}^3$ will be discussed further. Now, let us consider a space translation:

$$C' = \begin{pmatrix} 0 \\ k' \end{pmatrix},$$

the transformation matrix $P$ being the identity of $\mathbb{R}^4$. The transformation rule of the affine components of the torsor (2.3)$_2$ becomes:

$$J' = J + C'T^T - TC'^T.$$

It provides just the expected transport law (1.2) of the angular momentum:

$$l' = l + k' \times p,$$

while the linear momentum is unchanged. It is worthwhile to remark that the choice of contravariant indices for the torsor components is not casual but absolutely necessary for recovering the transport law (Souriau 1997a).

**Affine tensors**. We showed the pertinence of replacing in Mechanics the usual linear tensors by the affine tensors. Among the latter ones, the torsors are simply defined and have a very nice mechanical interpretation. The corresponding transformation law of their affine components is nothing else than the well-known transport law of the angular momentum. The previously defined types of affine tensors are summarized in Table 2.

*Table* 2. Affine tensors.

| Tensor type | space $AT_p^q \; {}^{contravariant}_{covariant} \}_{index}$ | abstract notations | transformation laws of the components |
|---|---|---|---|
| point of the affine space | $A\mathcal{T} = AT_0^1$ | $V$ | $V^{\alpha'} = C^{\alpha'} + (P^{-1})^{\alpha'}_{\beta} V^{\beta}$ |
| affine function $A\mathcal{T} \xrightarrow{\psi \atop \text{affine}} \mathbb{R}$ | $A^*\mathcal{T} = AT_1^0$ | $\psi$ | $\Phi_{\alpha'} = \Phi_{\beta} P^{\beta}_{\alpha'}$ <br> $\chi' = \chi - \Phi_{\beta} P^{\beta}_{\alpha'} C^{\alpha'}$ |
| torsor $A^*\mathcal{T} \times A^*\mathcal{T} \xrightarrow{\mu \; bilinear \atop skew-symmetric} \mathbb{R}$ | $AT_0^2$ | $\mu$ | $T^{\alpha'} = (P^{-1})^{\alpha'}_{\beta} T^{\beta}$ <br> $J^{\alpha'\beta'} = (P^{-1})^{\alpha'}_{\mu}(P^{-1})^{\beta'}_{\nu} J^{\mu\nu}$ <br> $+ C^{\alpha'}((P^{-1})^{\beta'}_{\mu} T^{\mu}) - ((P^{-1})^{\alpha'}_{\mu} T^{\mu}) C^{\beta'}$ |

**Using the linear representation of the affine group.** From the viewpoint of the calculus, it is sometimes useful to work with the linear representation of $\mathbb{GA}(n)$ in $\mathbb{R}^{n+1}$. The affine transformation $a = (C, P)$ is represented by the matrix:

$$\tilde{P} = \begin{pmatrix} 1 & 0 \\ C & P \end{pmatrix},$$

and its inverse is:

$$\tilde{P}^{-1} = \begin{pmatrix} 1 & 0 \\ C' & P^{-1} \end{pmatrix}.$$

Thus relation (2.1) can be written as:

$$\begin{pmatrix} 1 \\ V' \end{pmatrix} = \begin{pmatrix} 1 & 0 \\ -P^{-1}C & P^{-1} \end{pmatrix} \begin{pmatrix} 1 \\ V \end{pmatrix} = \begin{pmatrix} 1 & 0 \\ C' & P^{-1} \end{pmatrix} \begin{pmatrix} 1 \\ V \end{pmatrix}.$$

Formally, the transformation law for the affine components of a point of an affine space of dimension $n$ takes the form of the transformation law for the components of a vector of a linear space of dimension $(n+1)$:

$$\tilde{V}' = \tilde{P}^{-1} \tilde{V}.$$

Similarly, the components of an affine function follow the rule: $\tilde{\psi}' = \tilde{\psi} \tilde{P}$, that allows to recover (2.2) as:

$$(\chi' \quad \Phi') = (\chi \quad \Phi)\begin{pmatrix} 1 & 0 \\ C & P \end{pmatrix} = (\chi \quad \Phi)\begin{pmatrix} 1 & 0 \\ -PC' & P \end{pmatrix}.$$

Where $\Phi$ is a row vector collecting the $\Phi_{\alpha}$. The affine components of a torsor can be stored in a $(n+1) \times (n+1)$ skew-symmetric matrix:

$$\mu(\psi, \hat{\psi}) = \Phi J \hat{\Phi}^T + (\chi \hat{\Phi} - \hat{\chi} \Phi)T = (\chi \quad \Phi)\begin{pmatrix} 0 & T^T \\ -T & J \end{pmatrix}\begin{pmatrix} \hat{\chi} \\ \hat{\Phi}^T \end{pmatrix} = \tilde{\psi} \tilde{\mu} \tilde{\psi}^T. \quad (2.6)$$

Then the rule $\tilde{\mu}' = \tilde{P}^{-1} \tilde{\mu} \tilde{P}^{-T}$ allows recovering law (2.3).

**More about this topic.** In this short introduction, we briefly presented the notion of affine tensorial algebra strictly useful for applications to the mechanics that will be addressed in the sequel. For more detail on the affine dual space, affine tensor product, affine wedge product and affine tangent bundles, the reader interested in this topic is referred to the so-called AV-differential geometry of Tulczyjew et al. 1988, Grabowska et al. 2004.

# 3. Galilean torsors

**Galilei group**. All what has been said so far about the affine tensors may be applied as much for General Relativity as for Classical Mechanics. Henceforth we shall restrict the analysis to the latter theory. Let us consider Galilei group, a subgroup of the affine group $\mathbb{GA}(4)$, collecting the Galilean transformations, that is the affine transformations $a = (C, P)$ such that (Souriau 1997-b):

$$C = \begin{pmatrix} \tau \\ k \end{pmatrix}, \qquad P = \begin{pmatrix} 1 & 0 \\ u & R \end{pmatrix},$$

where $u \in \mathbb{R}^3$ is a Galilean boost, $R \in \mathbb{SO}(3)$ is a rotation, $k \in \mathbb{R}^3$ is a spatial translation and $\tau \in \mathbb{R}$ is a clock change. The inverse transformation is $a^{-1} = (C', P^{-1})$ given by:

$$C' = -P^{-1} C = \begin{pmatrix} \tau' \\ k' \end{pmatrix}, \qquad P^{-1} = \begin{pmatrix} 1 & 0 \\ -R^T u & R^T \end{pmatrix},$$

where $\tau' = -\tau$, $k' = R^T(-k + u\tau)$. Galilei group is a Lie group of dimension 10. Any Galilean transformations can be seen as a coordinate change: $X' = C' + P^{-1} X$, hence:

$$t' = t + \tau', \qquad r' = R^T(r - u t) + k'. \tag{3.1}$$

**Galilean torsors**. Latter on, the analysis is restricted to the class of the tensors the components of which are changed only by Galilean transformations. Naturally, we call them the Galilean tensors. The class of the Galilean tensors is a subclass of the class of affine tensors (as the class of Euclidean tensors is a subclass of the linear tensors). For Galilean torsors, the transformation rule (2.3) leads to:

$$m' = m, \tag{3.2}$$
$$p' = R^T(p - m u), \tag{3.3}$$
$$l' = R^T(l + u \times q) + k' \times (R^T(p - m u)), \tag{3.4}$$
$$q' = R^T(q - \tau'(p - m u)) + m k'. \tag{3.5}$$

**Galilean torsor invariants**. Here, we use standard methods of differential geometry to construct invariants (Olver 1995). The key idea is considering that the scalar $m$ and the components of the column-vectors $p, l, q$ can be collected to give a column-vector $\hat{\mu}$ of $\mathbb{R}^{10}$. The relations (3.2-5) show how Galilei group acts on the space $\mathbb{R}^{10}$, seen as a differential manifold of dimension 10 and called in the sequel torsor component space. We want now to determine a functional basis, *i.e.* a complete set of independent functions of these components, invariant under the group action.

The first step is to calculate their number. In the sequel, we will use some elementary tools of Lie group theory. For more details, the reader is referred for instance to Abraham and Marsden 1978. We start determining the isotropy group of $\hat{\mu}$, *i.e.* the set of Galilean transformations that let unchanged a given $\hat{\mu}$. The analysis will be restricted to massive particles: $m \neq 0$. The components $m, p, l, q$ being given, we have to solve the following system:

$$m = m, \tag{3.6}$$
$$p = R^T(p - m u), \tag{3.7}$$
$$l = R^T(l + u \times q) + k' \times (R^T(p - m u)), \tag{3.8}$$
$$q = R^T(q - \tau'(p - m u)) + m k', \tag{3.9}$$

with respect to $\tau', k', R, u$. Equation (3.6) is automatically satisfied (in fact, we identify a first invariant, the component $m$). Owing to (3.7), the boost $u$ can be expressed with respect to the rotation $R$ by:

$$u = \frac{1}{m}(p - R\,p).\qquad(3.10)$$

Next, owing to (3.7), equation (3.9) can be simplified as follows:

$$q = R^T q - \tau' p + m k',$$

that allows to determine the spatial translation $k'$ with respect to $R$ and the clock change $\tau'$:

$$k' = \frac{1}{m}(q - R^T q + \tau' p).\qquad(3.11)$$

Finally, because of (3.7), equation (3.8) is simplified as follows:

$$l = R^T(l + u \times q) + k' \times p.$$

Substituting (3.11) into the last relation gives:

$$l - \frac{1}{m} q \times p = R^T \left( l + u \times q - \frac{1}{m} q \times (R\,p) \right).$$

Then, (3.7) gives: $R\,p = p - m u$, that leads to:

$$l - \frac{1}{m} q \times p = R^T \left( l - \frac{1}{m} q \times p \right).$$

To simplify, let us introduce the column-vector:

$$l_0 = l - \frac{1}{m} q \times p.\qquad(3.12)$$

These quantity being given, we have to determine the rotations satisfying:

$$l_0 = R^T l_0.\qquad(3.13)$$

If $l_0$ does not vanish, the solutions of (3.13) are the rotations of an arbitrary angle $\theta$ about the axis $l_0$. We know by (3.10-11) that $u$ and $k'$ are determined in a unique manner with respect to $R$ and $\tau'$. The isotropy group of $\hat{\mu}$ can be parameterised by $\theta$ and $\tau'$. It is a Lie group of dimension 2, included in Galilei group, of which the dimension is 10. The orbit of $\hat{\mu}$ is the set of $\hat{\mu}'$ that can be obtained by applying a Galilean transformation to $\hat{\mu}$. It is a closed submanifold of the torsor component space. Its dimension is equal to the difference between the dimension of Galilei group and the one of the isotropy group: 10 – 2 = 8 (Abraham and Marsden 1978). The number of independent invariant functions is the difference between the dimension of the torsor component space and the one of the orbit: 10 – 8 = 2. A possible functional basis is composed of $m$ and the Euclidean norm $\|l_0\|$ of $l_0$. In the particular case $l_0 = 0$, all the rotations of $\mathbb{SO}(3)$ satisfy (3.13), then the isotropy group is of dimension 4. By similar reasoning to the case of non vanishing $l_0$, we conclude that the number of invariant functions is 4. A possible functional basis is composed of $m$ and the three null components of $l_0$.

**Physical interpretation of the Galilean torsor components**. Using (3.3-5), it can been verified that, under a Galilean transformation, the transformation rule of $l_0$, defined by (3.12), is given by:

$$l'_0 = R^T l_0.\qquad(3.14)$$

The previous developments suggest considering a system of space-time coordinates $X'$ such that: $p' = q' = 0$. Using definition (3.12) in this system, we have: $l' = l'_0$. Now, we claim that the components of the torsor in this system characterise a particle at rest with respect to this system. Its trajectory equation is: $r' = 0$. Next, let us consider another system of coordinates $X$, in uniform motion with respect to the former one with transport velocity $v$. To know the torsor components in the new system, let us consider the Galilean transformation:

$$\tau' = 0, \qquad k' = -r_0, \qquad R = I_3, \qquad u = v.$$

By virtue of (3.1), we know that: $r' = r - v t - r_0$. The trajectory equation in the new coordinate system is:

$$r = r_0 + v t.  \tag{3.15}$$

The transformation law (3.2-5) of the Galilean torsor components gives:

$$m' = m, \qquad p = m v, \qquad l = l'_0 + q \times v, \qquad q = m r_0.  \tag{3.16}$$

By (3.14), it holds: $l'_0 = l_0$. Combining (3.15) and (3.16), one obtains:

$$p = m v, \qquad l = l_0 + m r \times v, \qquad q = m (r - v t).  \tag{3.17}$$

The scalar $m$ is invariant, then characteristic of the particle and independent with respect to any coordinate system. It can be identified to the mass. The quantity $p$, product of the mass by the velocity, is the linear momentum. The column-vector $q$ is often hushed up in classical treatises, although it is not less basic than the first ones. Its meaning is simple. The column vector $q$ is the product of the mass by the initial position $r_0$ at time $t = 0$, which is obviously constant. The quantity $q$ is called passage by Souriau 1997-a, 1997-b, because the particle is passing through $r_0$ at time $t = 0$. Finally, $l$ is the angular momentum and $l_0$ is the spin (or intrinsic angular momentum).

In conclusion, the mass, the passage, the linear and angular momenta are structured as a single object, the torsor. Using the linear representation of an affine space, we can represent the torsor in a compact way thanks to the $(n+1) \times (n+1)$ skew-symmetric matrix used in (2.6):

$$\tilde{\mu} = \begin{pmatrix} 0 & m & p^T \\ -m & 0 & -q^T \\ -p & q & -j(l) \end{pmatrix}.$$

## 4. Affine Connection

**Affine tangent space.** In the previous section, the space-time was considered as a flat affine space, that was sufficient to describe the motion of free particles without interactions. With the aim of modelling free falling particles, we claim now that the space-time is a manifold $\mathcal{M}$ of dimension $n$. $T_X \mathcal{M}$ denotes the tangent space at $X$. While the difference of the components of two points of an affine space is defined without ambiguousness, the difference of the coordinates of two points in a manifold *has no meaning* in general. To get round this difficulty, the key idea is to consider that $A\mathcal{T}$ is the affine space associated to the tangent linear space $\mathcal{T} = T_X \mathcal{M}$. It is denoted $A T_X \mathcal{M}$ and called the *affine tangent space* at $X$. As observed by É. Cartan, 1923: "*The affine space at point **m** could be seen as the manifold itself that would be perceived in an affine manner by an observer located at **m***". The transformation matrices of Section 2 are just the Jacobean matrices of coordinate changes $(Y^{\beta'}) \mapsto (X^\alpha)$:

$$P^\alpha_{\beta'} = \frac{\partial X^\alpha}{\partial Y^{\beta'}} \ .$$

**Affine connection**. The manifold $\mathcal{M}$ is equipped with a symmetric linear connexion: $\omega^\beta_\alpha = \Gamma^\alpha_{\rho\beta} d X^\rho$, which specifies the infinitesimal motion of the current basis in the tangent space. This allows to define an intrinsic derivative of a vector field $X \mapsto \vec{V}(X) \in T_X \mathcal{M}$ :

$$\nabla V^\alpha = d V^\alpha + \omega_\beta^\alpha V^\beta . \tag{4.1}$$

To equip the manifold with an affine connexion, we need the previous connexion matrix $\omega_\beta^\alpha$ and an additional connection column vector specifying the motion of the current origin of the affine tangent space (de Saxcé et al. 2003):

$$\omega_C^\alpha = \Gamma_{\rho C}^\alpha d X^\rho = (\delta_\rho^\alpha - \nabla_\rho V_0^\alpha) d X^\rho . \tag{4.2}$$

We denote $\tilde{\nabla}$ the corresponding affine derivative symbol, in order to distinguish it from the one of the classic linear covariant derivative $\nabla$. The affine connexions were introduced by É. Cartan, 1923 and 1924. In de Saxcé et al. 2003, we show that the affine covariant derivative of a field of points of the affine tangent space $X \mapsto V(X) \in AT_X \mathcal{M}$ is given by:

$$\tilde{\nabla} V^\alpha = \nabla V^\alpha + \omega_C^\alpha = d V^\alpha + \omega_\beta^\alpha V^\beta + \omega_C^\alpha .$$

Next, we calculate the affine covariant derivative of the affine functions:

$$\tilde{\nabla} \Phi_\alpha = \nabla \Phi_\alpha , \qquad \tilde{\nabla} \chi = \nabla \chi - \Phi_\alpha \omega_C^\alpha .$$

In de Saxcé et al. 2003, for modelling the dynamics of shells, we consider a three-dimensional sub-manifold of the space-time and torsor fields with vector values in the tangent space to the sub-manifold. From the previous formula, we derived a general formula to calculate the affine covariant divergence of vector-valued torsor fields. For the dynamics of material particles, the sub-manifold is the trajectory, hence of dimension one. This is a particular case where the torsor is scalar-valued, as defined in Section 2. The general formula in de Saxcé et al. 2003 gives for the affine derivative of scalar-valued torsor fields:

$$\tilde{\nabla} T^\alpha = \nabla T^\alpha , \qquad \tilde{\nabla} J^{\alpha\beta} = \nabla J^{\alpha\beta} + \omega_C^\alpha T^\beta - T^\alpha \omega_C^\beta ,$$

or, in a matrix form :

$$\tilde{\nabla} T = \nabla T , \qquad \tilde{\nabla} J = \nabla J + \omega_C T^T - T \omega_C^T . \tag{4.3}$$

## 5. The motion equations of a free falling particle

**Galilean coordinate systems**. Any coordinate change representing a rigid body motion and a clock change:

$$r' = (R(t))^T (r - r_0(t)), \quad t' = t + \tau_0 ,$$

where $t \mapsto R(t) \in \mathbb{SO}(3)$ and $t \mapsto r_0(t) \in \mathbb{R}^3$ are smooth mappings, and $\tau_0 \in \mathbb{R}$ is a constant, is called a *Galilean coordinate change*. Indeed, a coordinate change is Galilean if and only if the corresponding Jacobean matrix is a linear Galilean transformation (Künzle 1972):

$$P^{-1} = \frac{\partial X'}{\partial X} = \begin{pmatrix} 1 & 0 \\ -R^T u & R^T \end{pmatrix},$$

where:

$$u = \varpi(t) \times (r - r_0(t)) + \dot{r}_0(t) , \tag{5.1}$$

is the well-known *velocity of transport*, with Poisson's vector $\varpi$ such that: $\dot{R} = j(\varpi) R$. There exists a family of coordinate systems which are deduced one from each other by such a coordinate changes. We call them *Galilean coordinate systems*.

**Galilean connections.** At each group of transformation is associated a family of connections and the corresponding geometry. We call Galilean connections the symmetric connections associated to Galilei group (Toupin 1958, Truesdell and Toupin, 1960, Künzle 1972). In a Galilean coordinate system, they are given by:

$$\omega = \begin{pmatrix} 0 & 0 \\ j(\Omega) d r - g d t & j(\Omega) d t \end{pmatrix}, \tag{5.2}$$

where $g$ is a column-vector collecting the $g^j = -\Gamma_{00}^j$ and identified to the gravity (Cartan, 1923), while $\Omega$ is a column-vector associated by the mapping $j^{-1}$ to the skew-symmetric matrix the elements of which are $\Omega_j^i = \Gamma_{j0}^i$. The vector $\Omega$ can be interpreted as representing Coriolis' effects (Souriau, 1997-a).

**Motion equations.** When the gravitation effects can be neglected, any particle is free and its motion is straight and uniform. The components of its torsor (mass, linear and angular momenta, passage) are time-independent or, in other words, constants of the motion. In a differential form, one reads:

$$d\mu = 0. \qquad (5.3)$$

When the particle is subjected to the gravitation effects, we say that it is a free falling particle. The geometrical structure of the space-time is a differential manifold. We generalize the previous equation in an intrinsic manner, and we claim that the motion equation can be written:

$$\tilde{\nabla}\mu = 0. \qquad (5.4)$$

In a detailed form:

$$\tilde{\nabla}T^\alpha = \nabla T^\alpha = 0, \qquad \tilde{\nabla}J^{\alpha\beta} = 0. \qquad (5.5)$$

**Linear momentum equation.** Now, let us examine the first equation. In a more explicit form, accounting for (4.1), one has:

$$\nabla T^\alpha = dT^\alpha + \omega_\beta^\alpha T^\beta = 0,$$

or, in an equivalent matrix form:

$$\nabla T = dT + \omega T = 0.$$

Let us examine in details these equations for Galilean connections. Owing to (2.4) and (5.2), it holds:

$$dm = 0, \qquad dp = mg\,dt - \Omega \times (m\,dr + p\,dt).$$

Dividing by $dt$ and accounting for $v = dr/dt$, one has:

$$\dot{m} = 0, \qquad \dot{p} = mg - \Omega \times (mv + p).$$

If we take into account $(3.17)_1$, one obtains:

$$\dot{m} = 0, \qquad \dot{p} = m(g - 2\Omega \times v). \qquad (5.6)$$

The first condition means the mass conservation along the trajectory. The second one defines the rate of linear momentum according to the connection. Among all the Galilean connections, there exists only one corresponding to our physical world. According to Newton theory, any particle of mass $m$, position $r$ at the time $t$, is subjected, in presence of another particle of mass $m'$, position $r'$ at the same time $t$, to a rate of linear momentum equal to:

$$\dot{p} = -\frac{k_g\,m\,m'}{\|r-r'\|^2}\frac{r-r'}{\|r-r'\|}, \qquad (5.7)$$

where $k_g$ is the constant of gravitation. The identification of (5.7), deduced from the experimental observations, with (5.6) suggests there exists particular Galilean coordinate systems for which ones:

$$g = -\frac{k_g\,m'}{\|r-r'\|^2}\frac{r-r'}{\|r-r'\|}, \qquad \Omega = 0.$$

The column vector $g$ is interpreted as the gravity due to the presence of other masses. The vector $\Omega$ is null in these particular Galilean coordinate systems (but not necessarily in other ones).

**Angular momentum equation.** Let us examine in details the consequences of the equation (5.5)$_2$. First of all, the *linear* covariant derivative (4.3)$_2$:

$$\nabla J = dJ + \omega J + J \omega^T,$$

of the angular momentum (2.5) has to be determined with respect to the Galilean connection (5.2):

$$\nabla J = d\begin{pmatrix} 0 & -q^T \\ q & -j(l) \end{pmatrix} + \begin{pmatrix} 0 & 0 \\ j(\Omega)dr - g\,dt & j(\Omega)dt \end{pmatrix}\begin{pmatrix} 0 & -q^T \\ q & -j(l) \end{pmatrix} + \begin{pmatrix} 0 & -q^T \\ q & j(l) \end{pmatrix}\begin{pmatrix} 0 & (j(\Omega)dr - g\,dt)^T \\ 0 & -j(\Omega)dt \end{pmatrix}.$$

With some abusive notations, we put:

$$\nabla J = \begin{pmatrix} 0 & -(\nabla q)^T \\ \nabla q & -j(\nabla l) \end{pmatrix}. \tag{5.8}$$

After several algebraic manipulations, it holds:

$$\nabla l = dl + \Omega \times l\,dt + q \times (\Omega \times dr - g\,dt), \tag{5.9}$$
$$\nabla q = dq + \Omega \times q\,dt, \tag{5.10}$$

In any Galilean coordinate system, we put for the affine connection:

$$\omega_A = \begin{pmatrix} d_A t \\ d_A r \end{pmatrix}. \tag{5.11}$$

Owing to (4.3)$_2$, (5.8) and (5.11), the affine covariant derivative of the angular momentum is given by:

$$\tilde{\nabla} J = \begin{pmatrix} 0 & -(\nabla q)^T \\ \nabla q & -j(\nabla l) \end{pmatrix} + \begin{pmatrix} d_A t \\ d_A r \end{pmatrix}\begin{pmatrix} m & p^T \end{pmatrix} - \begin{pmatrix} m \\ p \end{pmatrix}\begin{pmatrix} d_A t & (d_A r)^T \end{pmatrix}.$$

With some abusive notations again, we write:

$$\tilde{\nabla} J = \begin{pmatrix} 0 & -(\tilde{\nabla} q)^T \\ \tilde{\nabla} q & -j(\tilde{\nabla} l) \end{pmatrix}.$$

Finally, the affine covariant derivative of the angular momentum is given by:

$$\tilde{\nabla} l = \nabla l + d_A r \times p, \tag{5.12}$$
$$\tilde{\nabla} q = \nabla q + m\,d_A r - d_A t \cdot p. \tag{5.13}$$

Introducing (5.9-10) in (5.12-13), equation (5.5)$_2$ writes:

$$\boxed{\begin{aligned} dl + \Omega \times l\,dt + q \times (\Omega \times dr - g\,dt) + d_A r \times p = 0, & \qquad (5.14) \\ dq + \Omega \times q\,dt + m\,d_A r - d_A t \cdot p = 0. & \qquad (5.15) \end{aligned}}$$

Now, we wish to show these equations generalise a well know relation, the angular momentum law. Similarly to what be done at Section 3, we consider a Galilean coordinate system $(X'^\beta)$ in which the particle is at rest: $v' = 0$. Let $X \mapsto (\vec{e}_{\alpha'}(X))$ be the moving frame associated to these coordinates. Let us consider the corresponding proper affine frame $X \mapsto (V'_0(X), (\vec{e}_{\alpha'}(X))) = (0, (\vec{e}_{\alpha'}(X)))$. It was seen that the particle can be characterised by a Galilean torsor such that $p' = q' = 0$ and $l' = l'_0$.

With respect to any Galilean coordinate system $(X^\alpha)$, the particle has the position $r$ and a velocity:

$$v = \frac{dr}{dt}, \tag{5.16}$$

which, by virtue of the law of velocity composition, is equal to the transport velocity. This suggests considering the following affine transformation:

$$\tau' = 0, \qquad k' = -r, \qquad R = I_3, \qquad u = v.$$

Accounting for (3.14), the transformation rule (3.2-5) of torsor components gives:

$$m = m', \qquad p = m v, \qquad l = l_0 + q \times v, \qquad q = m r. \tag{5.17}$$

To calculate the affine covariant derivative of the angular momentum, we need to determine the affine connexion (4.2). In matrix form, relation (2.1) reads:

$$V_0' = C' + P^{-1} V_0. \tag{5.18}$$

The initial affine frame being proper, one has: $V_0' = 0$. By the affine transformation (5.18), we obtain:

$$V_0 = -P C' = -\begin{pmatrix} 1 & 0 \\ -v & I_3 \end{pmatrix} \begin{pmatrix} 0 \\ -r \end{pmatrix} = \begin{pmatrix} 0 \\ r \end{pmatrix}.$$

The affine connexion is calculated thanks to (4.2) that is written in a matrix form:

$$\omega_A = d X - \nabla V_0 = d X - d V_0 - \omega V_0.$$

Thus, one has:

$$\omega_A = \begin{pmatrix} d_A t \\ d_A r \end{pmatrix} = d \begin{pmatrix} t \\ r \end{pmatrix} - d \begin{pmatrix} 0 \\ r \end{pmatrix} - \begin{pmatrix} 0 & 0 \\ j(\Omega) d r - g d t & j(\Omega) d t \end{pmatrix} \begin{pmatrix} 0 \\ r \end{pmatrix} = \begin{pmatrix} d t \\ -\Omega \times r \, d t \end{pmatrix}. \tag{5.19}$$

Introducing expression (5.19) of the affine connexion in the angular momentum law (5.14-15), taking into account (5.16) and dividing by $d t$, one has:

$$\dot{l} + \Omega \times l + q \times (\Omega \times v - g) - (\Omega \times r) \times p = 0,$$

$$\dot{q} + \Omega \times (q - m r) - p = 0.$$

Applying Jacobi's identity, accounting for the definition (3.12) of the angular momentum and for (5.17), we obtain after simplification:

$$\dot{l} + \Omega \times l_0 = r \times m (g - 2 \Omega \times v), \tag{5.20}$$

$$\dot{q} = p. \tag{5.21}$$

The last relation is elegant. It can be understood by considering the conservation of the mass and the expression of $q$ and $p$ given by (5.17). In a Galilean coordinate system where $\Omega = 0$, the first relation is the usual expression of the law of the angular momentum:

$$\dot{l} = r \times m g.$$

The crucial point we would like emphasize is that (5.14-15) generalises this well-known law in a full covariant form, consistently with Galilei relativity principle. Besides, these equations can describe the motion of free falling particles but also the one of a rigid body, as it will be shown in the next section.

## 6. The motion equations of a rigid body

**Material and spatial representations.** To characterise the motion of a rigid body, we consider a Galilean coordinate system $(X^{\alpha'})$ in which one any point with a position $s'$ at the time $t'$ is at rest:

$$v' = \frac{d s'}{d t'} = 0. \tag{6.1}$$

The $s^{j'}$ are the classical Lagrangean or material coordinates of the point. The domain occupied by the rigid body is denoted $V$. The mass distribution is defined by a measure $s' \mapsto d m(s')$ on $V$. The total mass is given by:

$$m = \int_V dm.$$

We would like to represent the global motion of the body. If it is observed from a very far point with respect to its size, it becomes so small that it can be considered as equivalent to a particle. This scaling will permit to represent the body motion around its centre of mass thanks to the spin particle model as previously described. Let $(X^\alpha)$ be an arbitrary Galilean coordinate system in which a particle lies at position $x$ at the time $t$. The $x^i$ are the classical Eulerian or spatial coordinates. This representation is linked to the material one by a Galilean coordinate change:

$$x = r(t) + R(t) s', \qquad t = t'. \tag{6.2}$$

For convenience, $t \mapsto r(t)$ will give the position of the centre of mass:

$$m r = \int_V x \, dm. \tag{6.3}$$

Owing to (5.1) and (6.1), the particle of position $s'$ in the material representation has, in the spatial representation, the velocity:

$$v = \frac{dx}{dt} = u = \varpi(t) \times (x - r(t)) + \dot{r}(t). \tag{6.4}$$

**Calculating the torsor of the body.** Here, we follow a procedure already used for the study of shells (de Saxcé et al. 2002). Each particle of the body is supposed spin free. In a proper frame of $AT_X \mathcal{M}$, the spin of the body particles is null. At large scale, the body can be perceived as a particle of which we would like to determine, by an appropriate integration, the torsor resulting from all the contributions of the particles it is made of. The crucial point is concerned by the angular momentum. Before the integration, it must be determined, not with respect to $x$, but with respect to the centre of mass $r$.

If the curvature effects of the space-time are neglected at the body scale, the manifold $\mathcal{M}$ can be approximated by the affine tangent space $AT_X \mathcal{M}$ at the current point $X$. The event $X$ representing a particle with position $x$ at the time $t$ is identified with the origin of the proper frame, that is the zero vector of the linear tangent space $T_X \mathcal{M}$. The event $X_c$ representing the centre of mass, with position $r$ at the time $t$, is identified with a point of the affine tangent space $AT_X \mathcal{M}$. Its affine coordinates are:

$$V = \begin{pmatrix} t \\ r \end{pmatrix} - \begin{pmatrix} t \\ x \end{pmatrix} = \begin{pmatrix} 0 \\ r - x \end{pmatrix}.$$

Let us take this point as new origin. In the non proper frame $r' = (V, (\vec{e}_\alpha))$, the considered point has new coordinates: $V' = 0$. Using (2.1), we consider the affine transformation $a = (C', P^{-1})$ with $P = I_4$ and:

$$C' = V' - V = \begin{pmatrix} 0 \\ x - r \end{pmatrix}.$$

Owing to the transformation law (2.3), the components of the linear momentum are conserved while the ones of the angular momentum are not null anymore:

$$T' = T, \qquad J' = C' T^T - T C'^T. \tag{6.5}$$

The torsor components must be considered as measures. The linear momentum is given by:

$$T' = \begin{pmatrix} dm \\ u \, dm \end{pmatrix}.$$

For the convenience of notations, the prime will be omitted in the sequel. Taking into account (6.5), the angular momentum in the new frame is given by:

$$q = (x - r) dm, \qquad l = (x - r) \times u \, dm.$$

The torsor at $X$ is now defined by its components with respect to the origin representing in $AT_X \mathcal{M}$ the point $X_c$ associated to the center of mass. By integrating these components over the body, we obtain the angular momentum of the rigid body, perceived at large scale as a particle with spin:
$$q_0 = \int_V (x-r)\,dm, \qquad l_0 = \int_V (x-r)\times u\,dm.$$
Because of (6.3), the passage vanishes:
$$q_0 = 0, \qquad l_0 = \int_V (x-r)\times u\,dm.$$
Using (6.2-4) and the double cross product formula, we get:
$$l_0 = R\,\mathsf{J}_0\,R^T\,\varpi,$$
where occurs the inertia matrix in the material representation:
$$\mathsf{J}_0 = \int_V (\|s\|^2\,I_3 - s\,s^T)\,dm(s).$$

**Motion equations in spatial representation.** Let us consider a Galilean coordinate system for which Coriolis effects are null:
$$\Omega = 0.$$
Once again, we start from the general equation (5.14-15). In the considered proper frame, the linear momentum $p$ and the passage $q = q_0$ vanish, while the angular momentum $l$ is reduced to the spin $l_0$. The equation (5.14) degenerates into:
$$\dot{l}_0 = 0,$$
which provides three integrals of the motion:
$$R(t)\,\mathsf{J}_0\,(R(t))^T\,\varpi\,(t) = l_0 = C^{te}. \tag{6.6}$$

**Motion equations in material representation.** The connection matrix is not a tensor and is modified in a coordinate change following the law:
$$\omega' = P^{-1}(\omega\,P + d\,P),$$
that gives for the change between spatial and material representation:
$$\omega' = P^{-1}(\omega\,P + d\,P) = \begin{pmatrix} 1 & 0 \\ -R^T u & R^T \end{pmatrix}\left[\begin{pmatrix} 0 & 0 \\ -g\,dt & 0 \end{pmatrix}\begin{pmatrix} 1 & 0 \\ u & R \end{pmatrix} + \begin{pmatrix} 0 & 0 \\ d\,u & d\,R \end{pmatrix}\right] = \begin{pmatrix} 0 & 0 \\ R^T(d\,u - g\,dt) & R^T\,j(\varpi)\,R\,dt \end{pmatrix}$$
By identification with:
$$\omega' = \begin{pmatrix} 0 & 0 \\ j(\Omega')\,d\,s' - g'\,dt & j(\Omega')\,dt \end{pmatrix},$$
Coriolis' effect $\Omega'$ in the material representation is obtained with respect to Poisson's vector:
$$\Omega' = R^T\varpi. \tag{6.7}$$
General equation (5.14) particularized to the material representation with $p' = q' = 0$ but not vanishing $\Omega'$ and given by (6.7) leads to:
$$\dot{l}'_0 + \Omega'\times l'_0 = 0. \tag{6.8}$$
On the other hand, owing to (6.6-7), transformation rule (3.14) gives:
$$l'_0 = \mathsf{J}_0\,\Omega'.$$
Introducing this last expression in law (6.8), one has:
$$\mathsf{J}_0\,\dot{\Omega}' + \dot{\Omega}'\times(\mathsf{J}_0\,\dot{\Omega}') = 0.$$
Left multiplying each member by $\Omega'^T$ gives: $\Omega'.(\mathsf{J}_0\,\dot{\Omega}') = 0$, and:
$$\Omega'.(\mathsf{J}_0\,\Omega') = C^{te}$$
That leads to the construction of Poinsot inertia ellipsoid.

**Remark.** Of course, if the bodies are subjected to equations of constraint such that rolling contact, fixed points, pivot joints or hinges, the variables are not independent one of each other. In this case, the previous formulation could be modified by introducing additional force of constraints or Lagrange multipliers. These techniques are standard and are not recalled here (for instance, see J.-L. Synge 1960).

# 7. Conclusion and perspectives

In this paper, we showed the crucial role played by the affine group in Mechanics. We proposed an extension of the usual rules of the "Tensorial Algebra" able to define the torsors as affine tensors, which allows recovering the well-known transport law of the angular momentum as the transformation law for the affine components of the torsors. Some of the viewpoints and tools used here are closed to the ones presented in the AV-differential geometry approach (Tulczyjew et al. 1988, Grabowska et al. 2004).

In the second part, we presented a generalization of the usual "Tensorial Analysis" in order to differentiate in an intrinsic way the fields of affine tensors. Hence, the well-known theorem of the angular momentum is recast in a more general form which highlights the intrinsic form of this law with respect to any moving origin. This goal was reached without introducing, as in Tulczyjew 1985, additional dimension to the manifold.

The extension of this theoretical frame to the Mechanics of continua is obtained by considering vector valued torsors and a principle of free divergence torsor, revealing the affine covariance of the motion equations of such continua. The Dynamics of three dimensional bodies and shells was presented in de Saxcé et al. 2002. The case of strings, arches and beams is briefly presented in de Saxcé et al. 2005-b and will be detailed in a next paper.

As this paper was mainly devoted to the Mechanics, we limited the mathematical aspects to the strictly necessary notions. More pervasive investigation should invoke the theory of the connections on a G-principal bundle where G is the structural group. According to the considered application, G can be the linear group, the affine group, Galilei or Poincaré group, the projective group and so on. Early, the projective connection was proposed by Veblen et al. 1930 and Schouten 1935 with highlights to Kalusa-Klein version (Kaluza 1918) of the General Relativity. Also É. Cartan (1924) developed a theory of projective connections. Our opinion is that the affine connection seems a powerful tool for a unified approach of the Mechanics and a good compromise between the linear connections, widely used but scaled-down for physical applications, and the projective ones of which the relevancy is still today uncertain despite of the attention paid by scientists.

The principle of covariance is one of the fundamental concepts of the General Relativity (see for instance Misner et al. 1973). Nevertheless, it is restricted to the linear frames. Considering now the affine frames, the extension of the present work to the General Relativity is straightforward. In Einstein's theory, the metrics plays a crucial role because its components are the potentials of the gravitation field. This suggests to consider the Euclidean affine tensors for Poincaré group $\mathbb{ISO}(1,3)$ (a subgroup of the affine group $\mathbb{GL}(4)$). Besides, this circumstance is strongly related to the Symplectic Mechanics (Souriau, 1997). Hence, the torsors with mixed indices, covariant and contravariant, can be defined and putted into duality with the affine connections on the principal $\mathbb{ISO}(1,3)$-bundle of the orthonormal affine coframes, leaning to a nice extension of Kirillov-Kostant-Souriau theorem (Souriau 1997, Abraham et al. 1978), as we will show in a next paper.